\documentclass[aps,prl,reprint,superscriptaddress,amsmath,amssymb,preprintnumbers,showpacs]{revtex4-1}
\usepackage{bm}
\usepackage{pdfpages} 
\usepackage{pgffor}
\makeatletter
\AtBeginDocument{\let\LS@rot\@undefined}
\makeatother

\usepackage{enumerate}
\usepackage{graphicx}
\usepackage{color}
\usepackage{float} 
\usepackage{xspace}
\usepackage{bm} 
\graphicspath{{./}}

\newcommand{\ket}[1]{\bigl| #1 \bigr>} 
\newcommand{\bra}[1]{\bigl< #1 \bigr|} 
\newcommand{\op}[1]{\hat{#1}}

\newcommand{\uv}[1]{\ensuremath{\hat{\mathbf{#1}}}} 
\renewcommand{\vec}[1]{{\mathbf{\bm{#1}}}}

\begin{document}

\title{Signatures of multi-band effects in high-harmonic generation in monolayer MoS$_2$}
\author{Lun Yue}
\thanks{These authors contributed equally to the work.}
\affiliation{Department of Physics and Astronomy, Louisiana State University, Baton Rouge, Louisiana 70803, USA}
\author{Richard Hollinger$^*$}
\email{richard.hollinger1@gmail.com}
\author{Can B. Uzundal}
\author{Bailey Nebgen}
\affiliation{Department of Chemistry, University of California Berkeley, Berkeley, CA 94720, USA}
\affiliation{Lawrence Berkeley National Laboratory, Materials Sciences Division, Berkeley, CA 94720, USA}
\author{Ziyang Gan}
\author{Emad Najafidehaghani}
\author{Antony George}
\affiliation{Institute of Physical Chemistry, Friedrich Schiller University Jena, 07743 Jena, Germany}
\author{Christian Spielmann}
\affiliation{Institute of Optics and Quantum Electronics, Friedrich Schiller University Jena, 07743 Jena, Germany}
\affiliation{Abbe Center of Photonics, Friedrich Schiller University Jena, 07745 Jena, Germany}
\affiliation{Helmholtz Institute Jena, 07743 Jena, Germany}
\author{Daniil Kartashov}
\affiliation{Institute of Optics and Quantum Electronics, Friedrich Schiller University Jena, 07743 Jena, Germany}
\affiliation{Abbe Center of Photonics, Friedrich Schiller University Jena, 07745 Jena, Germany}
\author{Andrey Turchanin}
\affiliation{Institute of Physical Chemistry, Friedrich Schiller University Jena, 07743 Jena, Germany}
\affiliation{Abbe Center of Photonics, Friedrich Schiller University Jena, 07745 Jena, Germany}
\author{Diana Y. Qiu}
\affiliation{Department of Mechanical Engineering and Materials Science, Yale University, New Haven, Connecticut 06520, USA}
\author{Mette B. Gaarde}
\affiliation{Department of Physics and Astronomy, Louisiana State University, Baton Rouge, Louisiana 70803, USA}
\author{Michael Z\"urch}
\email{mwz@berkeley.edu}
\affiliation{Department of Chemistry, University of California Berkeley, Berkeley, CA 94720, USA}
\affiliation{Lawrence Berkeley National Laboratory, Materials Sciences Division, Berkeley, CA 94720, USA}
\affiliation{Institute of Optics and Quantum Electronics, Friedrich Schiller University Jena, 07743 Jena, Germany}


\begin{abstract}
    High-harmonic generation (HHG) in solids has been touted as a way to probe ultrafast dynamics and crystal symmetries in condensed matter systems. Here, we investigate the polarization properties of high-order harmonics generated in monolayer MoS$_2$, as a function of crystal orientation relative to the mid-infrared laser field polarization. At several different laser wavelengths we experimentally observe a prominent angular shift of the parallel-polarized odd harmonics for energies above approximately 3.5 eV, and our calculations indicate that this shift originates in subtle differences in the recombination dipole strengths involving multiple conduction bands. This observation is material specific and is in addition to the angular dependence imposed by the dynamical symmetry properties of the crystal interacting with the laser field, and may pave the way for probing the vectorial character of multi-band recombination dipoles. 
\end{abstract}

\maketitle



High-harmonic generation (HHG) from intense laser pulses in solids is increasingly attracting attention as a probe of ultrafast electronic processes in solids \cite{Vampa2017tutorial, Kruchinin2018review, Ghimire2019review}. HHG-spectroscopy in condensed matter has been used to investigate phase transitions \cite{Bionta2021PRRes} as well as quantum phenomena like the Berry curvature \cite{Liu2017NPhys, Luu2018NCommun}, electron correlation in the condensed phase \cite{Silva2018NPhoton, Murakami2021PRB, Orthodoxou2021npjQM, Jensen2021PRB}, and topological surface states \cite{Bai2021NPhys, Schmid2021Nature, Baykusheva2021PRA}. Solid-state HHG results from carrier dynamics following the excitation of electrons in the strong field of a laser pulse, and originates from both a nonlinear intraband current and, in case of recombination at high kinetic energies, an interband polarization \cite{Vampa2014PRL, Floss2018PRA, Yue2020PRL}. HHG from atomically thin semiconductors such as some of the transition metal dichalcogenides (TMDs) have gained special interest \cite{Liu2017NPhys, Yoshikawa2019NCommun, Cao2021OE, Liu2020NJP, Guan2020APL, Kobayashi2021US, Heide2021arxiv}. In addition to their interesting fundamental properties and their potential applications for optoelectronics and spintronics \cite{Manzeli2017review}, TMDs are ideal for investigating the microscopic origin of HHG due to the absence of laser propagation effects in the material \cite{Floss2018PRA, Kilen2020PRL, Hussain2021APL}.

The characteristics of the harmonic emission depend critically on the combined temporal translation symmetry of the laser field and the spatial symmetry of the system, referred to as the dynamical symmetry (DS) \cite{Neufeld2019NCommun}. An example well known from gas-phase HHG is the absence of the emission of even-order harmonics in centrosymmetric systems driven by linearly polarized lasers. In crystalline systems, the presence of multiple spatial symmetry elements have major consequences for the harmonic selection rules and polarization states \cite{Luu2016PRB, Klemke2019NCommun, Langer2017NPhoton, Jiang2019JPB}. For example, it was shown in the TMD MoS$_2$ that even-order harmonics are polarized perpendicularly when the driving laser is polarized along the zig-zag direction \cite{Yoshikawa2019NCommun, Liu2017NPhys}. In addition to the crystal symmetry properties, material-specific properties can also strongly affect harmonic characteristics, since different band structures and dispersion relations will give rise to different carrier dynamics. 
For example, it was recently demonstrated that the polarization properties of HHG can deviate in different TMDs \cite{Kobayashi2021US}, even though they in principle should obey the same dynamical symmetry selection rules. The understanding of the interplay between the symmetry- and material-specific properties in HHG is crucial to the probing of material-specific properties such as band structures \cite{Vampa2015PRL} and the vectorial character of the transition dipoles \cite{Uchida2021PRB}.
Accurate extraction of such material properties based on measurement of solid-state HHG spectra will allow leveraging of the available femtosecond time resolution of solid-state HHG spectroscopy to trace, e.g., phase transitions in materials \cite{Bionta2021PRRes}.

While much of the microscopic HHG dynamics can be understood in terms of a two-band picture \cite{Vampa2014PRL, Vampa2015Nature, Jiang2018PRL, Yoshikawa2019NCommun, Yue2020PRL, Kobayashi2021US, Uchida2021PRB} involving a filled valence band and a single conduction band, a number of works have highlighted the role of multiple bands \cite{Hawkins2015PRA, Wu2016PRA, Ikemachi2017PRA, Du2018PRA, Yu2020PRA, Liu2020NJP}. In particular, it has been experimentally demonstrated that multiple bands were responsible for nonintuitive harmonic waveforms in bulk GaSe \cite{Hohenleutner2015Nature}, the emergence of multiple plateaus in rare-gas solids \cite{Ndabashimiye2016Nature}, and yield enhancement of different harmonic regions in MgO \cite{Uzan2020NPhoton}. In this respect, the observation and understanding of multi-band effects in HHG from TMDs is of fundamental interest, and can potentially lead to the ultrafast probing of the complex dynamics leading to HHG, such as electron-hole recombinations from different valence and conduction bands, as well as using HHG to probe dynamic material properties that imprint at higher energies such as hot electron injection in a heterostructure.

In this letter, in a joint experimental and theoretical effort, we investigate the polarization properties of HHG in monolayer MoS$_2$, as a function of crystal orientation relative to the mid-infrared (mid-IR) laser polarization.
The DSs of the crystal combined with the temporal translation symmetries of the laser field imposes clear constraints on the harmonic emissions and leads to pronounced nodes in the HHG spectra at certain crystal orientations. In addition to the constraints and selection rules imposed by the DSs, we find an angular shift such that for several mid-IR wavelengths, the parallel-polarized odd-order harmonics below (above) 3.5 eV are enhanced for driver polarization along the armchair (zigzag) direction. By solving the time-dependent density matrix (TD-DM) equations involving a valence band and two conduction bands, we trace these angular shifts to electron-hole recombinations from different conduction bands, effectively probing the vectorial nature of recombination dipoles in different bands. Our work reveals that electron-hole recombinations from different conduction bands are directly imprinted on the orientation-dependent HHG spectra, and opens up the possibility of probing electron-hole dynamics from multiple recombination pathways.


HHG experiments were performed with mid-IR laser pulses having wavelength between 3 and 5 $\mu$m with full polarization control reaching 1.5 TW/cm$^2$ in the focus. All experiments were performed at ambient conditions.
Monolayer single crystals of MoS$_2$ were grown on SiO$_2$/Si substrate by chemical vapor deposition (CVD) \cite{George2019JPM, Shree20192DM} and transferred on to a double side polished c-plane oriented sapphire substrate using a PMMA assisted transfer protocol \cite{George2019JPM}. The MoS$_2$ monolayer crystals were characterized by optical microscopy, atomic force microscopy (AFM) and Raman spectroscopy.
More details on the experimental setup is given in the Supplemental Material (SM) \cite{suppmat2022_mos2_multiband}.
The 1H-semiconducting phase of MoS$_2$ sample was confirmed by investigating the HHG signal from a circular polarized laser and finding that only the $3n\pm1$ harmonic orders were generated \cite{Jia2020PRB}.


\begin{figure}
  \centering
  \includegraphics[width=0.48\textwidth, clip, trim=0 0cm 0 0cm]{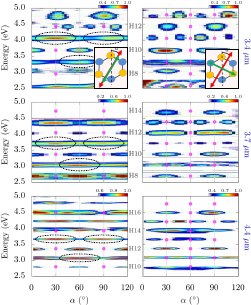}
  \caption{Experimental HHG spectra measured in the parallel (left column) and perpendicular configuration (right column) for MoS$_2$ irradiated by intense mid-IR pulses of central wavelength 3.4 $\mu$m (top row), 3.7 $\mu$m (middle row) and 4.4 $\mu$m (bottom row). Yields are plotted in linear scale and arbitrary units. In the insets, the two polarizer/analyzer configurations are sketched, with $\alpha$ being the angle between the crystal mirror plane and the laser polarization (red arrow), and the polarizer direction indicated by the green arrow. The magenta bullets mark the nodes predicted by a DS analysis, and the dashed ellipses mark the angular shifts of the parallel odd harmonics.}
  \label{fig:tmdc_2}
\end{figure}

We show the measured harmonic emission from monolayer MoS$_2$ in Fig.~\ref{fig:tmdc_2}, using three different fundamental laser wavelengths: 3.4, 3.7 and 4.4 $\mu$m. The configurations in which we detect harmonics (green arrows) polarized parallel or perpendicular to the laser polarization (red arrows) are sketched in the insets. We denote $\alpha$ as the angle between the laser polarization and the crystal mirror plane, and for notational convenience label the $\Gamma-M$ (armchair) direction as $\alpha_{0} \equiv n 60^\circ$ and the $\Gamma-K$ (zigzag) direction as $\alpha_{30}\equiv 30^\circ+n60^\circ$, with $n$ being an integer. In Fig.~\ref{fig:tmdc_2}, the detected harmonic yields in the two configurations (columns) for different wavelengths (rows) are plotted as a function of harmonic energy and $\alpha$.

Figure~\ref{fig:tmdc_2} illustrates in detail the interplay between DS constraints and material-specific properties and how they manifest on the harmonic emission. 
There are three notable features:
First, the DS constraints give rise to characteristic nodes (magenta bullets) in the angular dependence of both the parallel- and the perpendicular-polarized harmonics, which we will discuss in more detail below.
Second, the odd harmonics in the parallel configuration exhibit a characteristic shift in their angular dependence (angular shift), in which their yields peak at $\alpha_0$ ($\alpha_{30}$) for energies below (above) $\sim$3.5 eV. This shift is marked by the dashed ellipses. 
Third, we find additional minima in the angular dependence of the high-energy ($\sim$ 4.5 eV), even harmonics in the parallel configuration for 3.4 $\mu$m and 4.4 $\mu$m that are not imposed by the DS constraints. 

We begin with an analysis of the DSs, which consist of combinations of time and spatial symmetries that leave the time-dependent Hamiltonian invariant \cite{Neufeld2019NCommun}.
For the spatial symmetries, we can consider the spatial point group $C_{3v}$ instead of the full MoS$_2$ group $D_{3h}$, since the laser field is polarized in the MoS$_2$ (001) plane. We consider linearly polarized monochromatic laser fields that satisfy the time symmetry $\op{\tau}_2\vec{F}(t)=-\vec{F}(t)$, where $\op{\tau}_2$ is an operator that performs a half-cycle time-translation. For a laser polarized along $\alpha_0$, taken as the $y$-axis for convenience, the reflection $\hat{\sigma}^y$ in the $y$-axis is a DS operation. The harmonic yield is contained in the Fourier components $\vec{r}_n$ of the time-dependent polarization $\vec{r}(t) \equiv \bra{\psi(t)} \vec{r} \ket{\psi(t)}$. The invariance of $\vec{r}(t)$ under the DS operation \cite{Neufeld2019NCommun, Baykusheva2021PRA} leads to the constraints $ (-r_{n,x}, r_{n,y}) = (r_{n,x}, r_{n,y})$, which gives the selection rule that perpendicular harmonics are forbidden along $\alpha_0$. Similarly, for a laser polarized along $\alpha_{30}$, taken now as the $x$-axis for convenience, the reflection together with the half-cycle translation of the pulse,  $\hat{\sigma}^y\op{\tau}_2$, constitutes a DS, which leads to the constraints $ (-r_{n,x}, r_{n,y}) e^{in\pi} = (r_{n,x}, r_{n,y})$ implying the selection rules that parallel even harmonics and perpendicular odd harmonics are forbidden for $\alpha_{30}$.

The predictions from the DS analysis in Fig.~\ref{fig:tmdc_2} 
are in clear agreement with the experimental results for all driver wavelengths (to within our signal-noise ratio), illustrating the strength of a general DS analysis and how it may enable the probing of the crystal structure for a known laser field, or the characterization of the field for a known crystal structure. Notably, the nodes implied by DS agree with the perpendicular odd harmonics having the unintuitive $30^\circ$-periodicity that at first glance deviates from the hexagonal lattice structure of MoS$_2$. Note that a similar $30^\circ$-periodicity was recently experimentally observed in GaSe \cite{Kaneshima2018PRL} and the topological insulator BSTS \cite{Bai2021NPhys}, with the origin explained purely in terms of the intraband harmonic generation mechanism. This explanation, however, cannot be applied here since our observed above-gap harmonics in MoS$_2$ (gap 1.8 eV) are predominantly generated by the interband mechanism \cite{Yoshikawa2019NCommun, Cao2021OE}.
Finally, as mentioned above, the results in Fig.~\ref{fig:tmdc_2} also demonstrate features that go beyond the DS symmetry, such as the angular shift, which we now will discuss. 


We first note that the angular shifts in Fig.~\ref{fig:tmdc_2} manifest near the same harmonic energy for all three laser wavelengths, strongly indicating that they are due to material-specific properties such as the band structure and dipole matrix elements.
To shed light onto the origin of these shifts, we proceed by solving the TD-DM equations in the velocity gauge and Bloch basis. We employ a linearly polarized field with cosine-squared envelope, full width at half maximum of 99 fs, wavelength 3.7 $\mu$m, and laser intensity of 0.15 TW/cm$^2$ [corresponding to a vector potential peak $A_0=0.168$ atomic units (a.u.)]
\footnote{In calculations, this intensity yields the best match to the experimental harmonic cutoff energies, and is consistent with a reduction of the electric field amplitude at the air-MoS$_2$ interface in combination with focal-volume averaging.}
During the time propagation, we transform to a basis that is dressed by the instantaneous field, to obtain the dressed density matrix elements $\bar{\rho}_{mn}^{\vec{k}}(t)$ and the dressed momentum matrix elements $\bar{\vec{p}}_{mn}^{\vec{k}}(t)$, where the subscripts refer to the band indices and $\vec{k}$ is the crystal momentum.
 This procedure allows us to include a phenomenological dephasing time (taken as 10 fs) and to decompose the current into an intraband and interband part. We note that this yields results in agreement with those of the popular multi-band semiconductor Bloch equations \cite{Golde2008PRB, Schubert2014NPhoton}, but has the advantage that a periodic structure gauge \cite{Yue2020PRA, Virk2007PRB} is not required. Including dephasing cleans up the harmonic spectra to a certain degree \cite{Floss2018PRA} but does not influence the conclusions drawn here. To model monolayer MoS$_2$, we employ the tight-binding model of Ref.~\cite{Liu2013PRB, Liu2020NJP}, which takes into account the Mo $d_{z^2}$, $d_{xy}$, and $d_{x^2-y^2}$ orbitals with third-nearest hoppings, and results in one valence band (VB) and two conduction bands (CB1 and CB2). The momentum matrix elements are calculated following Ref.~\cite{Pedersen2001PRB}. For more details on the theory, see the SM \cite{suppmat2022_mos2_multiband}.

\begin{figure}
  \centering
  \includegraphics[width=0.48\textwidth, clip, trim=0 0cm 0 0cm]{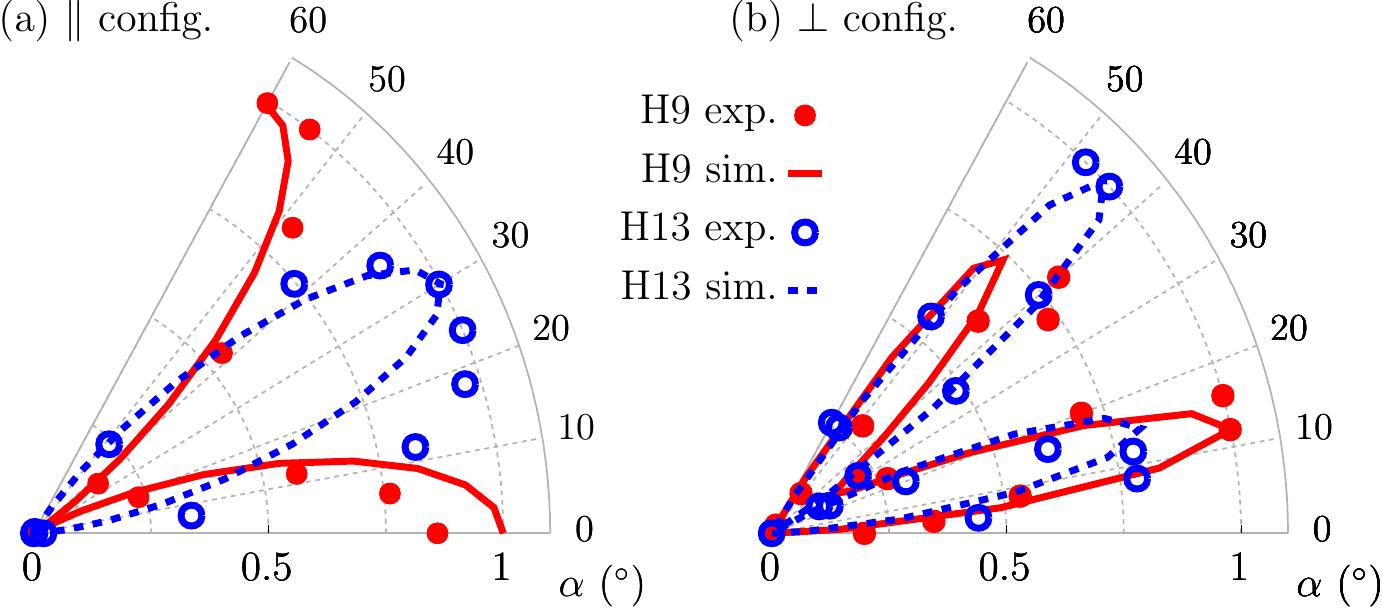}
  \caption{Experimental and simulation $\alpha$-dependent HHG yields with a 3.7 $\mu$m driver, for the (a) parallel and (b) perpendicular configurations.
    The zero (unity) of the radial axes is set as the yield minimum (maximum).}
  \label{fig:tmdc_3}
\end{figure}

Figure~\ref{fig:tmdc_3} depicts a comparison of measured and calculated angular yields for harmonic 9 (H9) and H13 in the parallel [Fig.~\ref{fig:tmdc_3}(a)] and perpendicular [Fig.~\ref{fig:tmdc_3}(b)] configurations for the 3.7 $\mu$m driver. The simulations reproduce the angular shift in the parallel configuration, as well as the 30$^\circ$-periodicity in the perpendicular direction. As discussed above, the $30^\circ$-periodicity can be understood from the DS analysis. However, the DSs do not restrict the parallel odd harmonics, such that emissions for all polarization angles $\alpha$ are in principle allowed. We now investigate the observed angular shift.


The H9 and H13 are emitted above the band edge of MoS$_2$ ($\sim$H5) and originate in the interband currents. The interband harmonic emission process can be interpreted in terms of the recollision model \cite{Vampa2014PRL, Yue2021PRA}: 
first, an electron-hole pair is created by tunnel excitation of an electron from VB to CB1 near the minimum band gap at $K$; the electron-hole pair is propagated in their respective bands following the acceleration theorem $\vec{k}+\vec{A}(t)$ \cite{Bloch1929ZPhys}, and during propagation, the coupling between CB1 and CB2 can lead to electron excitation to CB2 \cite{Uzan2020NPhoton}; finally, the electron and hole can coherently recombine either from CB1 or CB2 to the VB, with a probability amplitude weighted by the recombination dipoles, and the harmonic energy given by the band gap at the time of recombination.

We consider the three steps of the recollision process, and discuss the differences between $\alpha_0$ and $\alpha_{30}$. First, it should be noted that the populations in the conduction bands at the end of the calculations are very similar: $1.33\%$ in CB1 and $0.84\%$ in CB2 for $\alpha_0$; and $1.32\%$ in CB1 and $0.86\%$ in CB2 for $\alpha_{30}$. This means that the initial excitation step cannot be responsible for the HHG yields between $\alpha_0$ and $\alpha_{30}$ in Figs.~\ref{fig:tmdc_3}(a) and \ref{fig:tmdc_3}(b). Since the band dispersions near the $K$ valley in the range $K\pm A_0$ are similar for the two conduction bands, the propagation step appears to play a minor role. This suggests that the angular shift is dominated by the recombination dipole which can be modeled further.

\begin{figure}
  \centering
  \includegraphics[width=0.48\textwidth, clip, trim=0 0cm 0 0cm]{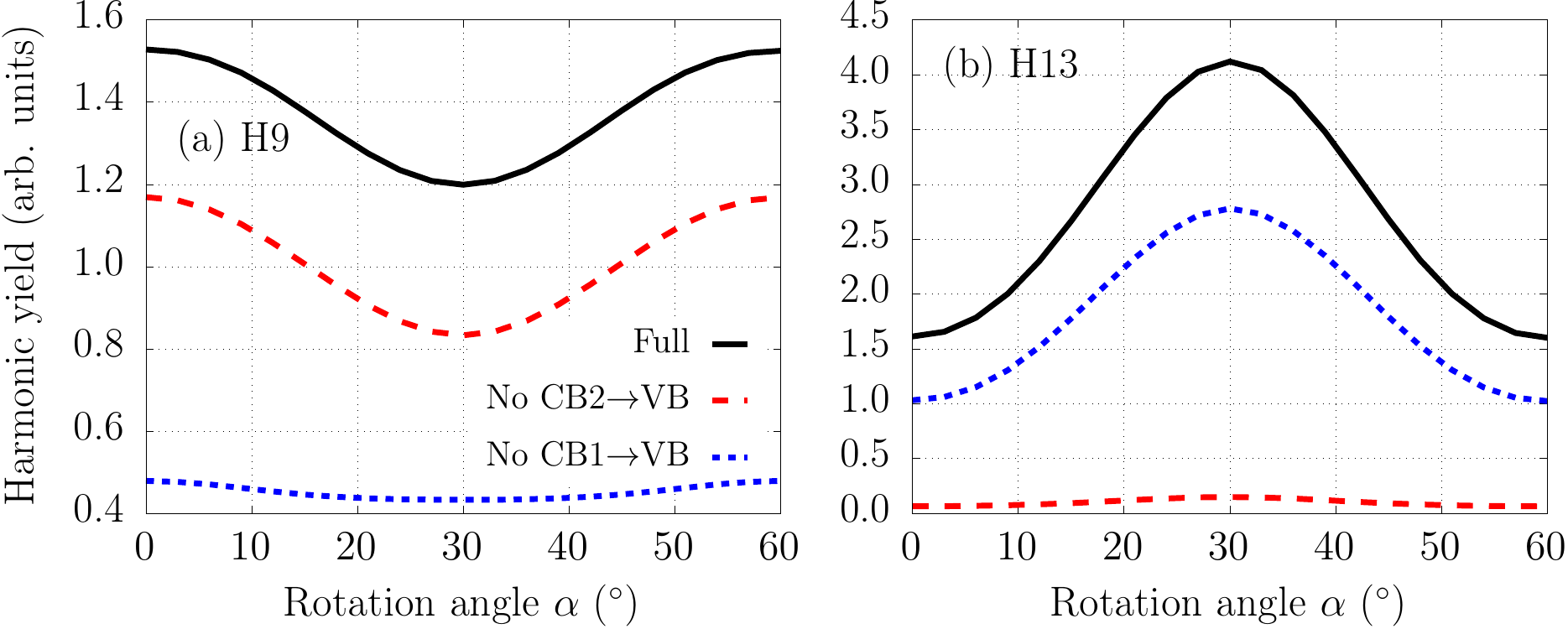}
  \caption{Simulated harmonic yields as a function of $\alpha$, for the parallel-polarized (a) H9 and (b) H13. The full (black) curves are the full calculations, the dashed (red) curves are for calculations with $\bar{\vec{p}}_{13}^{\vec{k}} = \vec{0}$, and the dotted (blue) curves are with $\bar{\vec{p}}_{12}^{\vec{k}} = \vec{0}$.}
  \label{fig:tmdc2d_hhyield}
\end{figure}

The full black curves in Fig.~\ref{fig:tmdc2d_hhyield} show the full simulations results, where the parallel polarized H9 in Fig.~\ref{fig:tmdc2d_hhyield}(a) peaks at $\alpha_0$ and H13 in Fig.~\ref{fig:tmdc2d_hhyield}(b) peaks at $\alpha_{30}$.
The dashed (red) curves are results using $\bar{\vec{p}}_{13}^{\vec{k}}(t)=\vec{0}$ in the calculations of the interband current, i.e. restricting the CB2$\rightarrow$VB recombination pathway, resulting in a yield almost exclusively from the CB1$\rightarrow$VB pathway. Similarly, the dotted (blue) curves are with $\bar{\vec{p}}_{12}^{\vec{k}}(t)=\vec{0}$, corresponding to the CB2$\rightarrow$VB recombinations. The H9 yield in Fig.~\ref{fig:tmdc2d_hhyield}(a) is seen to be dominated by recombination from CB1, while the H13 yield in Fig.~\ref{fig:tmdc2d_hhyield}(b) is dominated by recombinations from CB2. Clearly, the $\alpha$-dependence of the yield originate in different characters of of the recombination dipoles $\bar{\vec{p}}_{12}^{\vec{k}}(t)$ and $\bar{\vec{p}}_{13}^{\vec{k}}(t)$, which we now investigate further.

\begin{figure}
  \centering
 \includegraphics[width=0.48\textwidth, clip, trim=0 0cm 0 0cm]{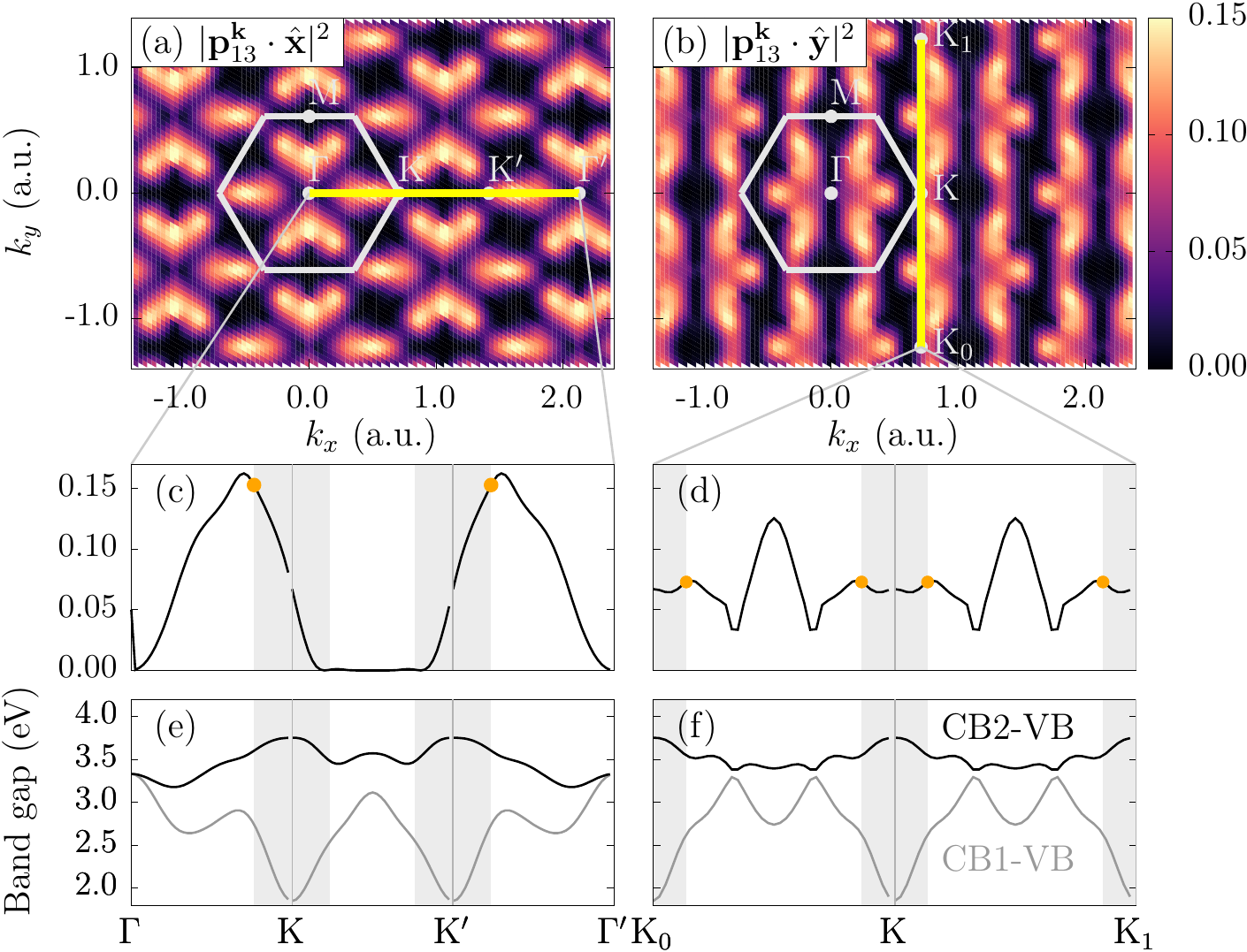}
  \caption{Modulus squares of the recombination dipoles along (a) $\uv{x}$ ($\alpha=30^\circ$) and (b) $\uv{y}$ ($\alpha=0^\circ$) directions. The gray hexagon traces the 1st Brillouin zone. Panels (c) and (d) plot the same as (a) and (b), but along paths in reciprocal space marked by the yellow lines. Panel (e) and (f) show the band gaps along these paths, with the gray (black) curve between CB1 (CB2) and the VB. The areas marked by gray indicate regions accessible by electron-hole pairs created at the $K$ point, with maximum excursions $\pm A_0$, and the orange dots in (c) and (d) highlight the maximum values in these regions.}
  \label{fig:tmdc2d_pmes}
\end{figure}

We choose to focus on the pronounced case, i.e. the bell shape of H13 in Fig.~\ref{fig:tmdc2d_hhyield}(b) with maximum at $\alpha_{30}$ and minimum at $\alpha_0$ dominated by the CB2$\rightarrow$VB pathway. The recombination matrix elements responsible for the parallel harmonics are proportional to $\vec{p}_{13}^{\vec{k}}\cdot \uv{e}_\alpha$, where $\uv{e}_\alpha$ is an unit vector along $\alpha$. In Fig.~\ref{fig:tmdc2d_pmes}, we plot $|\vec{p}_{13}^{\vec{k}}\cdot \uv{e}_\alpha|^2$ for $\alpha_{30}$ ($\uv{e}_\alpha\equiv \uv{x}$) in (a) and for $\alpha_{0}$ ($\uv{e}_\alpha \equiv \uv{y}$) in (b). The values along relevant paths in reciprocal space, indicated by the yellow lineouts in Figs.~\ref{fig:tmdc2d_pmes}(a) and \ref{fig:tmdc2d_pmes}(b), are shown in Figs.~\ref{fig:tmdc2d_pmes}(c) and Fig.~\ref{fig:tmdc2d_pmes}(d), respectively. An electron-hole pair created at $K$ at the zeros of the vector potential will have the maximal excursion $K\pm A_0 \uv{e}_\alpha$ in reciprocal space, which is marked by the gray regions in Figs.~\ref{fig:tmdc2d_pmes}(a) and Fig.~\ref{fig:tmdc2d_pmes}(b). The maxima of $|R_{13}^{\vec{k}}|^2$ in the gray regions are shown by the orange dots, with the values $0.153$ for $\alpha_{30}$ and $0.073$ for $\alpha_0$. The ratios between them is 2.10, which is qualitatively in agreement with 2.55 -- the ratio between the yields at $\alpha_{30}$ and $\alpha_0$ for the black curve in Fig.~\ref{fig:tmdc2d_hhyield}(b).
This suggests that the angular shift is due to the structure of recombination dipoles between the CBs and the VB, with the CB1$\rightarrow$VB recombination dominating the H9 and the CB2$\rightarrow$VB recombination dominating the H13. These considerations are further supported by considering the $\vec{k}$-dependent band gaps in Figs.~\ref{fig:tmdc2d_pmes}(e) and (f), where it is observed that the recombination energy from the CB2 to the VB in the allowed regions are above $\sim 3.5$ eV, in good agreement with the experimental observation (dashed ellipses in Fig.~\ref{fig:tmdc_2}) that the angular shift occur at around $3.5$ eV. We have thus established that the HHG spectrum carries direct information on the vectorial character of the recombination dipoles, in addition to the band structure information. We note that for the highest-order parallel even harmonics at 4.5 eV in Fig.~\ref{fig:tmdc_2} for 3.4 $\mu$m and 4.4 $\mu$m, the minima at $\alpha=60^\circ$ cannot be explained in terms of the DS or our three-band model, and we attribute these to interference effects from even higher conduction bands that are beyond the scope of this work.

Our combined experimental observations and theoretical analysis indicate the contributions of higher conduction bands to the HHG spectra from monolayer MoS$_2$ that goes beyond pure considerations of DS properties of the crystal and laser field. 
This was achieved by experimentally measuring the harmonics polarized parallel and perpendicular to the driving laser polarization for three different wavelengths, aided by careful theoretical analysis that selects the contributions from different recombination dipoles. The experimental signature manifests in the parallel-polarized harmonics below (above) 3.5 eV being favorably emitted when the laser polarization is along the armchair (zigzag) direction, and is observed at all three wavelengths used in the experiment.
Our work elucidates the novel interplay between the DSs of the system and material-specific properties.
We believe that our results may enable HHG spectroscopy of not only the band structure \cite{Vampa2015PRL}, but also the vectorial nature of the transition dipoles \cite{Uchida2021PRB} beyond a two-band description, as well as electron-hole recombinations from different conduction bands in other materials. Our work will also aid in the fundamental understanding of light-matter interactions in TMDs such as cycle-resolved currents and valleytronics \cite{Langer2018Nature, Hashmi2021arxiv}.

\begin{acknowledgments}
L. Y. and M. B. G. acknowledge support from the National Science Foundation, under Grants No. PHY-1713671 and PHY-2110317, and high performance computational resources provided by the Louisiana Optical Network Infrastructure (http://www.loni.org).
R. H. acknowledges support by the Alexander von Humboldt Foundation. B. N. acknowledges support by the National Science Foundation Graduate Research Fellowship Program. M. Z. acknowledges support by the Federal Ministry of Education and Research (BMBF) under ``Make our Planet Great Again - German Research Initiative'' (Grant No. 57427209 ``QUESTforENERGY'') implemented by DAAD. M. Z. acknowledges funding by the W. M. Keck Foundation, funding from the UC Office of the President within the Multicampus Research Programs and Initiatives (M21PL3263), and funding from Laboratory Directed Research and Development Program at Berkeley Lab (107573).
Z. G., E. N., A. G. and A. T. acknowledge support by DFG Collaborative Research Center SFB 1375 'NOA' Project B2, and thank Stephanie H\"oppener and Ulrich S. Schubert for enabling our Raman spectroscopy study at the JCSM.
C. S. and D. K. aknowledge support by CRC 1375 ``NOA-Nonlinear optics down to atomic scales'' funded by the Deutsche Forschungsgemeinschaft (DFG). 
D. Y. Q. was supported by the U.S. Department of Energy, Office of Science, Office of Basic Energy Sciences Early Career Research Program under Award Number DE-SC0021965.
\end{acknowledgments}

%




\section*{Supplementary material}

# Supplemental Material: Signatures of multi-band effects in high-harmonic generation in monolayer MoS$_2$

Lun Yue,[1,][*] Richard Hollinger[*],[2, 3,][†] Can B. Uzundal,[2, 3] Bailey Nebgen,[2, 3] Ziyang Gan,[4] Emad Najafidehaghani,[4] Antony George,[4] Christian Spielmann,[5, 6, 7] Daniil Kartashov,[5, 6] Andrey Turchanin,[4, 6] Diana Y. Qiu,[8] Mette B. Gaarde,[1] and Michael Zürch[2, 3, 5,][‡]

[1]*Department of Physics and Astronomy,*
*Louisiana State University, Baton Rouge, Louisiana 70803, USA*
[2]*Department of Chemistry, University of California Berkeley, Berkeley, CA 94720, USA*
[3]*Lawrence Berkeley National Laboratory,*
*Materials Sciences Division, Berkeley, CA 94720, USA*
[4]*Institute of Physical Chemistry, Friedrich Schiller University Jena, 07743 Jena, Germany*
[5]*Institute of Optics and Quantum Electronics,*
*Friedrich Schiller University Jena, 07743 Jena, Germany*
[6]*Abbe Center of Photonics, Friedrich Schiller University Jena, 07745 Jena, Germany*
[7]*Helmholtz Institute Jena, 07743 Jena, Germany*
[8]*Department of Mechanical Engineering and Materials Science,*
*Yale University, New Haven, Connecticut 06520, USA*

---

[*] These authors contributed to the work.
[†] richard.hollinger1@gmail.com
[‡] mwz@berkeley.edu

## I. EXPERIMENTAL MATERIAL CHARACTERIZATION

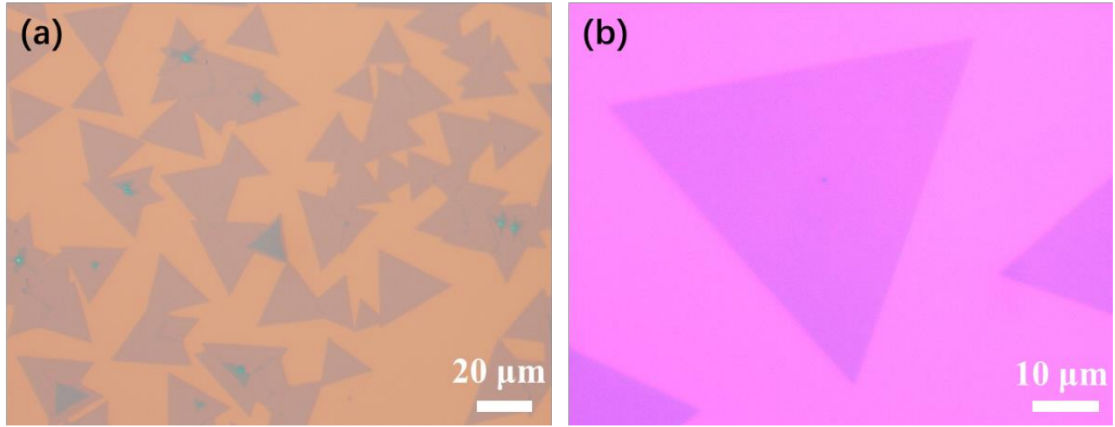

FIG. S1: (a)-(b) Optical microscopy images of monolayer $MoS_2$ single crystals grown on $SiO_2$/Si substrates by chemical vapor deposition (CVD) as described in Ref. [1]. The OM images were acquired with a Zeiss Axio Imager Z1.m microscope equipped with a 5-megapixel CCD camera (AxioCam ICc5) in bright field operation.

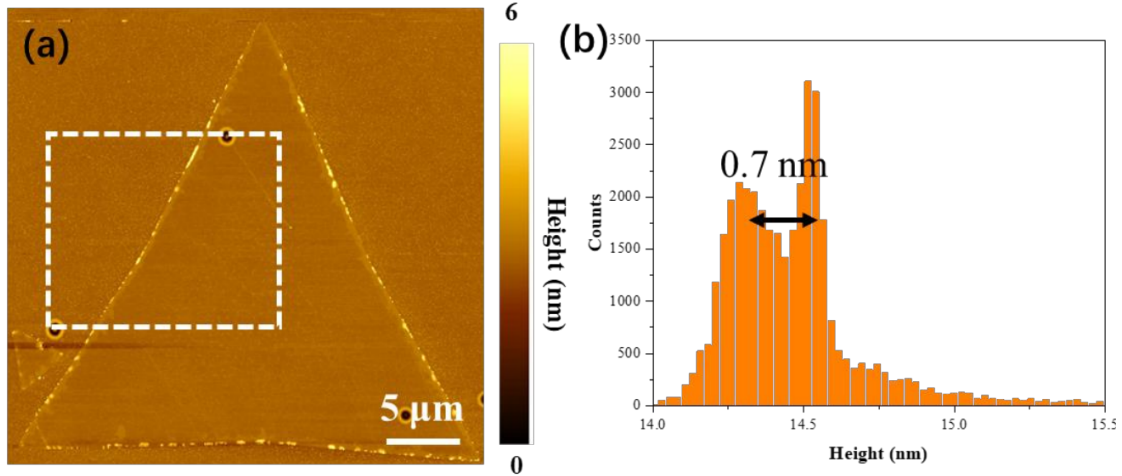

FIG. S2: (a) Atomic force microscopy (AFM) topography image of CVD grown $MoS_2$ monolayer on $SiO_2$/Si substrate. (b) Height distribution histogram between the $SiO_2$ surface and the surface of $MoS_2$. The thickness of the $MoS_2$ crystal is estimated as 0.7±0.2 nm, which corresponds to a monolayer. The AFM measurement was performed with a Ntegra (NT-MDT) system in the tapping mode at ambient conditions using n-doped silicon cantilevers (NSG01, NTMDT) with the resonant frequencies of 87-230 kHz and a typical tip radius of < 6 nm.

2

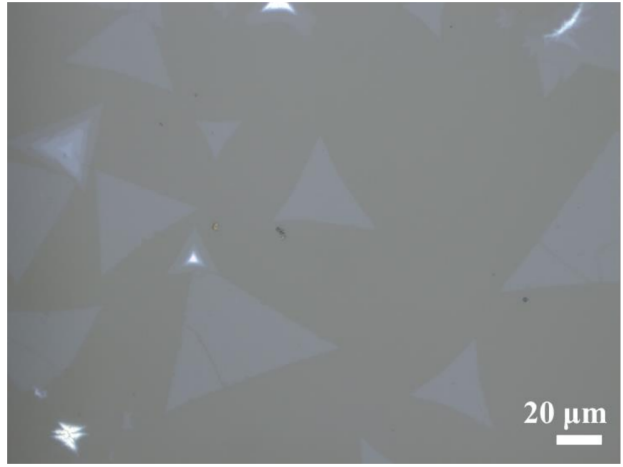

FIG. S3: Optical microscopy image of CVD grown monolayer $MoS_2$ crystals transferred onto c-plane oriented double side polished sapphire substrate.

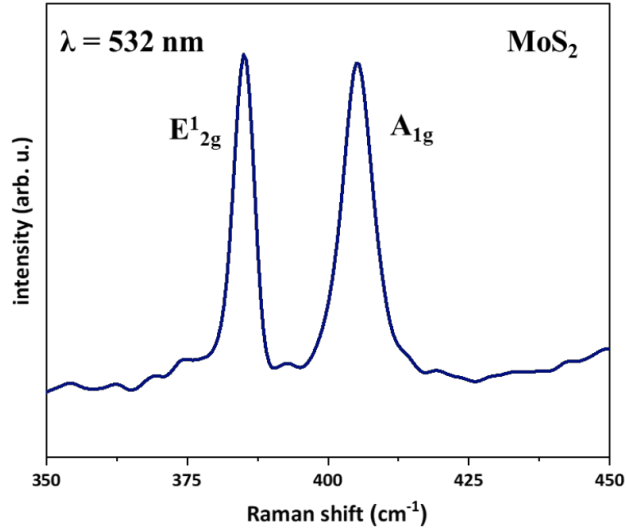

FIG. S4: Typical Raman spectrum recorded on CVD grown monolayer $MoS_2$ single crystals. Two characteristic peaks of $MoS_2$ are $E^1_{2g}$ at 385.5 cm$^{-1}$ originating from in-plane vibrations of Mo-S bonds and $A_{1g}$ originating from out-of-plane vibrations of S atoms at 404 cm$^{-1}$ are observed. The difference between the two peak positions is 18.5 cm$^{-1}$, which confirms the formation of monolayer single crystals [2]. The Raman spectra were acquired with a Bruker Senterra spectrometer operated in the backscattering mode at ambient conditions. The measurements was caried out using a 532 nm, frequency-doubled Nd:YAG Laser, a 50x objective and a thermoelectrically cooled CCD detector. The spectral resolution of the system is 2-3 cm$^{-1}$.

## II. EXPERIMENTAL SETUP FOR HHG

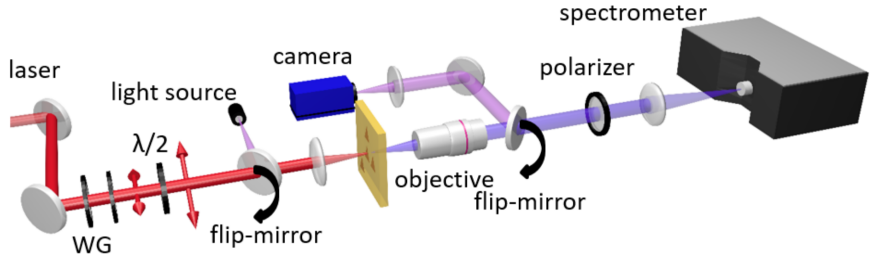

FIG. S5: Experimental setup for HHG in $MoS_2$ from a linear polarized field as a function of the crystal orientation. The mid-IR laser source in this experiment is tuned between 3 and 5 $\mu$m with full polarization control. Two wire grid (WG) polarizers enable intensity control.

HHG experiments were performed with mid-IR laser pulses, resulting from non-collinear difference frequency mixing of the signal and idler from a Ti:Sa-amplifier-pumped optical parametric amplifier. The mid-IR laser pulses with a repetition rate of 1 kHz had a central wavelength between 3 and 5 $\mu$m and a duration of 110 fs. As shown in Fig. S5, the laser power was attenuated using a pair of wire grid polarizers. The relative orientation between the laser polarization and the $MoS_2$ crystal was controlled by a broadband $\lambda/2$ plate. Using a wire grid attenuator the mid-IR intensity was adjusted below the damage threshold of the sample. The laser was focused down to a spot size radius of 55 $\mu$m achieving a vacuum intensity of 1.5 TW/cm$^2$. We analyzed the harmonic polarization component parallel and perpendicular to the laser linear polarization using a broadband wire grid polarizer. The harmonic signal was detected by a spectrograph equipped with a Peltier-cooled CCD. The monolayer sample used in the experiment was identified using an optical microscope setup consisting of a white light source and a CCD camera. All experiments were performed at room temperature and ambient conditions.

## III. SIMULATION DETAILS

### A. Band structure and momentum matrix elements

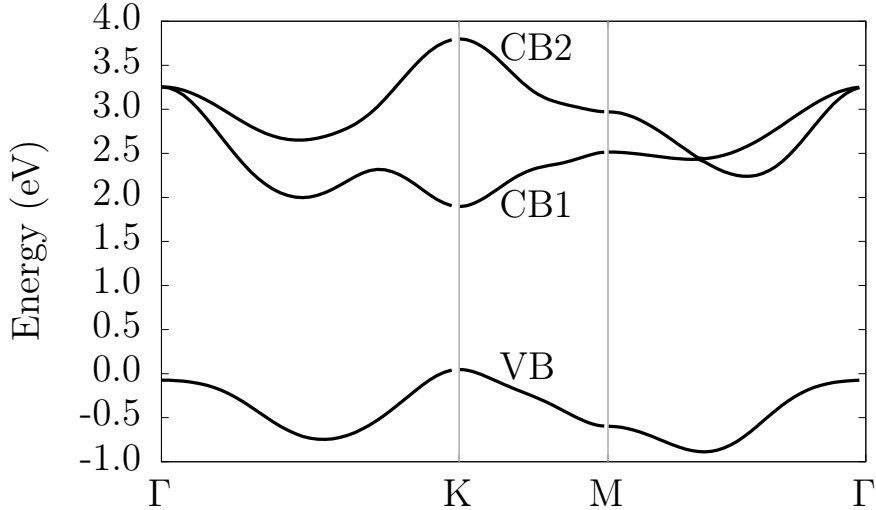

FIG. S6: MoS$_2$ band structure considered with one VB and two CBs.

The band structure of MoS$_2$ in the three-band model [3] is shown in Fig. S6. We calculate the momentum matrix elements as [4]

$$\mathbf{p}^{\mathbf{k}}_{mn} = \sum_{\alpha\beta} c^{\mathbf{k}*}_{\alpha m} c^{\mathbf{k}}_{\beta n} \nabla_{\mathbf{k}} H^{\mathbf{k}}_{\alpha\beta}, \quad (1)$$

where $H^{\mathbf{k}}_{\alpha\beta}$ are the tight-binding Hamiltonian matrix elements [3] and $c^{\mathbf{k}}_{\alpha m}$ are elements of the unitary matrix that diagonalizes the tight-binding Hamiltonian. In writing Eq. (1), we have neglected terms that contain the intra-atomic dipoles, as they are zero in the atomic $d$-orbital basis. Note also that for a laser field polarized in the xy-plane (monolayer plane), due to the reflection symmetry [3] in this plane, the conduction bands with $d_{xz}$ or $d_{yz}$ characters can never be populated - the two conduction bands considered in this work are thus sufficient for our purposes, especially given that the dynamics considered here occur near the $K$-valleys. The Brillouin zone is sampled as a Monkhorst-Pack mesh with $100 \times 100$ points.

5

## B. Time-dependent density-matrix propagation

The TD-DM equations in the velocity and Bloch basis read [5]

$$i\dot{\rho}^{\mathbf{k}}_{mn}(t) = \left(E^{\mathbf{k}}_m - E^{\mathbf{k}}_n\right)\rho^{\mathbf{k}}_{mn}(t) + \mathbf{A}(t) \cdot \sum_l \left[\mathbf{p}^{\mathbf{k}}_{ml}\rho^{\mathbf{k}}_{ln}(t) - \mathbf{p}^{\mathbf{k}}_{ln}\rho^{\mathbf{k}}_{ml}(t)\right], \quad (2)$$

with $\mathbf{k}$ the crystal momentum, $\rho^{\mathbf{k}}_{mn}$ the density matrix elements, $E^{\mathbf{k}}_m$ the band energies, and $\mathbf{p}^{\mathbf{k}}_{mn}$ the momentum matrix elements. The vector potential is taken as $\mathbf{A}(t) = A_0 \cos^2[\pi t/(2\tau)]\sin(\omega_0 t)\hat{\mathbf{e}}_\|$, with $t \in [-\tau, \tau]$, $\omega_0$ the carrier frequency, and the full width at half maximum of the field. The propagation time step is $\delta t = 0.2$, and for every 5th time step ($\Delta t = 1$) the density matrix is transformed into the basis that is instantaneously dressed with the field, $\rho^{\mathbf{k}}_{mn} \to \bar{\rho}^{\mathbf{k}}_{mn}$, where a phenomelogical dephasing $T_2 = 10$ fs is applied as $\bar{\rho}^{\mathbf{k}}_{mn} \to \bar{\rho}^{\mathbf{k}}_{mn} e^{-\Delta t/T_2}$ for $m \neq n$. Note that direct inclusion of a phenomenological dephasing term in the TD-DM of Eq. (2) would give unphysical results, due to the strong state-mixing. More details on our method will be presented in a future work.

The time-dependent current is calculated as the sum of an intraband and interband contribution, $\mathbf{j}(t) = \mathbf{j}^{tra}(t) + \mathbf{j}^{ter}(t)$, with

$$\mathbf{j}^{tra}(t) = \sum_{\mathbf{k}} \sum_m \bar{\mathbf{p}}^{\mathbf{k}}_{mm}(t)\bar{\rho}^{\mathbf{k}}_{mm}(t) \quad (3a)$$

$$\mathbf{j}^{ter}(t) = \sum_{\mathbf{k}} \sum_{m,n\neq m} \bar{\mathbf{p}}^{\mathbf{k}}_{mn}(t)\bar{\rho}^{\mathbf{k}}_{nm}(t), \quad (3b)$$

where $\bar{\mathbf{p}}^{\mathbf{k}}_{mn}(t)$ is the momentum matrix transformed to the dressed basis. The HHG spectrum polarized parallelly and perpendicularly to the laser polarization are calculated as $S_\|(\omega) \propto \left|\mathbf{j}(\omega) \cdot \hat{\mathbf{e}}_\|\right|^2$ and $S_\perp(\omega) \propto \left|\mathbf{j}(\omega) \cdot \hat{\mathbf{e}}_\perp\right|^2$, with $\mathbf{j}(\omega)$ the Fourier transform of the current weighted by a window function and $\hat{\mathbf{e}}_\perp$ a unit vector perpendicular to the laser polarization.

---




\end{document}